\magnification=\magstep1
\baselineskip=20 truept
\def\ni{\noindent}

\def\bn{\bigskip\noindent}
\def\mn{\medskip\noindent}

\def\ie{{\it i.e.}}
\def\pa{\partial}

\def\half{{\textstyle{1\over2}}}
\def\quar{{\textstyle{1\over4}}}
\def\rt{{\bf R}^3}
\def\ii{{\rm i}}  

\def\vp{\varphi}
\def\ve{\varepsilon}
\font\scap=cmcsc10
\font\title=cmbx10 scaled\magstep1

\def\l{\lambda}

\rightline{DTP-98/55}
\vskip 1truein
\centerline{\title Hopf Solitons on S$^3$ and R$^3$.}

\vskip 1truein
\centerline{\scap R. S. Ward}

\bn\centerline{\it Dept of Mathematical Sciences, University of Durham,}
\centerline{\it Durham DH1 3LE, UK.}

\vskip 2truein
\ni{\bf Abstract.} The Skyrme-Faddeev system, a modified O(3) sigma
model in three
space dimensions, admits topological solitons with nonzero Hopf number.
One may learn something about these solitons  by considering the system
on the 3-sphere of radius $R$.  In particular, the Hopf map is a solution
which is unstable if $R > \sqrt{2}$.

\vskip 1truecm
\ni PACS 11.27.+d, 11.10.Lm, 03.50.-z

\vfil\eject

\ni1. {\scap Introduction}

\mn The simplest (3+1)-dimensional system admitting localized topological
solitons is the O(3) nonlinear sigma model, in which
a configuration is a map $\vp:\rt\to S^2$.  Given
a suitable boundary condition at spatial infinity, $\vp$ extends to a map
from $S^3$ to $S^2$; and such maps are classified by their Hopf number
$Q\in \pi_3(S^2)\cong{\bf Z}$.  This raises the possibility of string-like
topological solitons, which can be linked or knotted; and interest in such
solitons has recently been rekindled by numerical investigations [1--5].  

To define the $\sigma$-model, one needs to choose a Lagrangian.  Let us
concentrate here on static fields only: so one has to choose an energy
functional $E[\vp]$ for maps  $\vp:\rt\to S^2$.  The simplest choice,
namely $E_2[\vp] = \int(\pa\vp)^2$, does not admit static solitons (as
the usual Derrick-Hobart scaling argument readily shows).  In order to have
stable solitons, one needs to prevent them from shrinking.  There are various
ways of achieving this; the one currently of most interest is the Skyrme
method, namely adding a functional $E_4[\vp]$ which is fourth-order in
derivatives of $\vp$.   In this paper, we deal with the resulting
Skyrme-Faddeev model [6]; so the energy functional is of the form
$E = E_2 + E_4$.

Over the years, there have been several attempts to understand soliton
solutions in this system.  The task is hampered by the fact that the
obvious ans\"atze are incompatible with the equations of motion [7].
Related to this is the fact that topologically nontrivial configurations
can admit at most an axial (one-parameter) symmetry [8].  So the situation
is unlike that of the Skyrme model, with target space $S^3$, where the
$Q=1$ soliton is a spherically-symmetric ``hedgehog''.  In the Skyrme-Faddeev
system, the minimal-energy $Q=1$ soliton is believed to be toroidal in shape.
Advances in
computer power have now made numerical investigations feasible, and the
recent work [1--5] involves numerical searches for static solitons.

In the case of the Skyrme model, one gains considerable insight from studying
the system on the 3-sphere of radius $R$, rather than on flat 3-space $\rt$
[9--11].  If $R=1$, then the single Skyrmion is simply the identity map
from $S^3$ to $S^3$, and it saturates the topological lower bound on the
energy.  As $R$ is increased, one reaches a point (in fact at $R = \sqrt{2}$)
when this identity map becomes unstable, and there is a spontaneous breakdown
of its symmetry.  For $R > \sqrt{2}$, and in the flat-space limit $R\to\infty$,
the single Skyrmion becomes localized around a particular point in space.

The purpose of this paper is to show that analogous results hold for Hopf
solitons in the Skyrme-Faddeev model.  In particular, we shall see that
the standard Hopf map from $S^3_R$ to $S^2$ is unstable if $R>\sqrt{2}$;
and we shall discuss the possibility of improving the topological lower
bound on the energy, so that it can be attained if $R=1$.

\bn 2. {\scap The Energy and its Topological Lower Bound}

\mn The system involves a scalar field $\vp$ which takes values on the
unit 2-sphere $S^2$.  We are interested here in static configurations only:
so  $\vp$ is defined on a positive-definite 3-space $(M, g_{jk})$, which
we shall take to be either $\rt$ (with standard Euclidean metric) or
$S^3_R$ (the standard 3-sphere of radius $R$).  In other words, we have a
nonlinear O(3) $\sigma$-model on $M$.  In the usual manner,  $\vp$ can be
written as a unit 3-vector field  $\vp^a = \vp^a(x^j)$, with
$\vp^a \vp^a = 1$.  [The indices $a,b,\ldots$ and  $j,k,\ldots$ run over
1,2,3;  the $x^j$ are local coordinates on $M$;  and the Einstein summation
convention applies throughout.]

The energy of  $\vp$ is defined by the functional
$$
  E := \int_M (c_2 {\cal E}_2 + c_4 {\cal E}_4) \, dV.
$$
Here $dV$ is the volume element $dV := \sqrt{g}\,dx^1\wedge dx^2\wedge dx^3$,
where $g = \det(g_{jk})$;  $c_2$ and $c_4$ are coupling constants;
and
$$\eqalign{
  {\cal E}_2 &:= g^{jk} (\pa_j \vp^a) (\pa_k \vp^a), \cr
  {\cal E}_4 &:= \quar g^{jl} g^{km} F_{jk} F_{lm}, \cr
}$$
where $F_{jk} := \ve_{abc} \vp^a (\pa_j \vp^b) (\pa_k \vp^c)$.  This is the
Skyrme-Faddeev system.  Note that  $c_2$ and $c_4$ are dimensional quantities,
with $\sqrt{c_4/c_2}$ having units of length and  $\sqrt{c_4 c_2}$ having
units of energy.

If $M = \rt$, then we impose the boundary condition that $\vp$ is smoothly
defined on the one-point compactification of $\rt$ (in particular, $\vp^a$
tends to a constant unit vector at spatial infinity).  So for the purposes
of differential topology, we may regard $\vp$ as a smooth map from $S^3$
to $S^2$.  Such maps are classified by an integer $Q$, namely the Hopf
number.  There is no local formula for $Q$ in terms of the field $\vp$;
a nonlocal expression is as follows.  Find $A_j$ such that
$F_{jk} = \pa_j A_k - \pa_k A_j$; this can be done because the 2-form
$F_{jk} dx^j\wedge dx^k$ is closed, and the deRahm cohomology group $H^2(S^3)$
is trivial.  Then
$$
  Q = {1\over32\pi^2} \int_M \eta^{jkl} F_{jk} A_l \, dV,
$$
where $\eta^{jkl} = g^{-1/2} \ve^{jkl}$ ($\ve^{jkl}$ being the totally-skew
symbol with $\ve^{123} = 1$).  The precise sign of $Q$, which has to
do with orientations, need not concern us; in what follows, we take $Q\geq0$.

It has been known for some time that there is a topological lower bound 
on the energy of configurations with $Q\neq0$, at least on $M = \rt$ [12].
The argument consists of a number of parts.  First, one has a Sobolev-type
inequality
$$
\biggl({1\over32\pi^2} \int_M \eta^{jkl} F_{jk} A_l \, dV\biggr)^{3/2}
 \leq C \biggl(\int_M {\cal E}_4 \, dV\biggr)
   \biggl(\int_M \sqrt{{\cal E}_4} \, dV\biggr), \eqno(1)
$$
where $C$ is a (universal) constant.  In other words, the integral of the
Chern-Simons form is bounded above by an expression involving integrals
of $F^2$ and $\sqrt{F^2}$.  A value for $C$ for which this
inequality holds, for $M = \rt$, is [12, 8]
$$
  C = {1 \over 8\sqrt{2}\, \pi^4\, 3^{3/4}}. \eqno(2)
$$
The second ingredient is the algebraic inequality
$$
   {\cal E}_4 \leq {1\over8}  {\cal E}_2^2 .
$$
This holds on $S^3_R$ as well as on $\rt$.  One way to establish it is to
consider the $3\times3$ matrix $D^{ab} = g^{jk} (\pa_j\vp^a)(\pa_k\vp^b)$
(cf.\ [10]).  This matrix has a zero eigenvalue (corresponding to the
eigenvector $\vp^a$); if we call the other two eigenvalues $\l_1$ and $\l_2$,
then ${\cal E}_2 =  \l_1 + \l_2$ and  ${\cal E}_4 =  \half \l_1 \l_2$.  The
inequality follows immediately.

The final ingredient is the simplest, namely
$$
  E \geq 2 \biggl(\int_M c_2 {\cal E}_2 \, dV\biggr)^{1/2}
   \biggl(\int_M c_4 {\cal E}_4 \, dV\biggr)^{1/2}.
$$
Putting the three inequalities together gives
$$
  E \geq K \, Q^{3/4}, \eqno(3)
$$
where $K = 2^{7/4}\, C^{-1/2} \,(c_2 c_4)^{1/2}$.

It seems likely that the value of $C$ given by (2) can be improved (\ie\
reduced).  The conjecture here , which is motivated by the Hopf map (see
section 3), is that (1) is true for
$$
  C = {1 \over 64\sqrt{2}\, \pi^4}. \eqno(4)
$$
It would follow that the value of $K$ in (3) is
$K_0 = 32\, \pi^2 \, (c_2 c_4)^{1/2}$; and this ``reference'' value is used
in the
discussion that follows.  Note that the value of $K$ obtained from (2)
is about half of $K_0$.

The recent numerical investigations have provided some data concerning
stable solutions (minimum-energy configurations) in the $Q=1$, $Q=2$
and $Q=3$ sectors on $M = \rt$.  One of these [2] assumed axial symmetry,
and obtained energies
$$\eqalign{
  E_{Q=1} &= 1.25\, K_0, \cr
  E_{Q=2} &= 1.19\, K_0\, 2^{3/4}. \cr
}$$
The other [3] was a fully three-dimensional simulation, and this yielded
$$\eqalign{
  E_{Q=1} &= 1.13\, K_0, \cr
  E_{Q=2} &= 1.11\, K_0\, 2^{3/4}, \cr
  E_{Q=3} &= 1.14\, K_0\, 3^{3/4}, \cr
  E_{Q=4} &= 1.18\, K_0\, 4^{3/4}, \cr
          &\vdots \cr
  E_{Q=8} &= 1.15\, K_0\, 8^{3/4}. \cr
}$$
The latter figures probably underestimate the true energies, owing to the
effect of working in a finite-volume subset of $\rt$.  So the current
evidence suggests that solitons have an energy around 20\% above our
conjectured lower bound.

\bn 3. {\scap Stability of the Hopf Map.}

\mn This section deals with the standard Hopf map from $S^3_R$ to $S^2$,
which has topological charge $Q=1$.  In particular, its energy is
$E = 16\pi^2\,(R+R^{-1})$, which is bounded below by $32\pi^2$ (this bound
being attained when $R=1$).  For  $R > \sqrt{2}$, by contrast, we shall see
that the Hopf map is unstable.  (From now on, $c_2$ and $c_4$
are set equal to 1; this sets the length and energy scales.)  These results are
analogous to those for Skyrmions [10].

The Hopf map $\Phi$ may be described as follows. A point of $S^3_R$
corresponds to a pair $(Z^0,Z^1)$ of complex numbers satisfying
$Z^A\bar Z_A = 1$, where $\bar Z_0 = \overline{Z^0}$ and 
$\bar Z_1 = \overline{Z^1}$.  The Hopf map sends this point to the point
on $CP^1 = S^2$ which has homogeneous coordinates  $[Z^0,Z^1]$.  The energy
density of this map $\Phi$ is constant on $S^3_R$; in fact it has
${\cal E}_2 = 8/R^2$ and
${\cal E}_4 = 8/R^4$, so that $E = 16\pi^2\,(R+R^{-1})$ as stated above.

The gauge potential $A^j$ turns out to be a right-invariant vector field
on $S^3$; in other words, it corresponds to (a generator of) $SU(2)$ acting on
$SU(2) \cong S^3$ by left multiplication.  The inequality (1) is a
statement about vector fields on $S^3$ (note that each side is independent
of the radius $R$).  For right-invariant vector fields, one has
equality in (1) with $C$ given by (4); and this suggests that (1) holds with
that that value of $C$.  But proving this would seem to require some
delicate global analysis, which will not be attempted here.

Each of $\int{\cal E}_2\, dV$ and  $\int{\cal E}_4\, dV$ is stationary
with respect to variations of the field: in other words, $\Phi$ is a
solution of the Euler-Lagrange equations for all values of $R$.  Clearly
one expects
this solution to be unstable for large $R$, and we shall now see this
in more detail.

The energy of a perturbation $\Phi + \delta\Phi$ of the Hopf map has the form
$$
  E[\Phi + \delta\Phi] =  E[\Phi] + \int_M G[\delta\Phi]\, dV,
$$
where $G[\delta\Phi]$ is a quadratic function of  $\delta\Phi$ and
$\pa_j(\delta\Phi)$.  We are interested in the eigenvalues (particularly
the negative eigenvalues) of the quadratic form
$\delta E = \int G[\delta\Phi]\, dV$.

Let us think of $Z^A$ as transforming under the fundamental representation of
$U(2)$.  The Hopf map is invariant under this $U(2)$, in the sense that
$\Phi(Z^A)$ and $\Phi(U^A_B Z^B)$ are related by an $SO(3)$ transformation
on the target space $S^2$.  Consequently, $U(2)$ also acts on the tangent
space at $\Phi$, \ie\ on the space of perturbations about  $\Phi$.  So we
can decompose this perturbation space into irreducible representations
of $U(2)$.

Hence we consider perturbations as follows: the perturbed field maps $Z^A$ to
the point on $CP^1$ with homogeneous coordinates  $Z^A + \Theta^A$, where
$$
  \Theta^A = T^{AB\ldots D}_{P\ldots R} \bar Z_B \ldots \bar Z_D
                          Z^P \ldots Z^R
$$
($T^{A\ldots}_{P\ldots}$ being a constant infinitesimal tensor).  In terms
of the unit 3-vector field $\vp^a(Z^B, \bar Z_B)$, the Hopf map is given by
$$\eqalign{
  \vp^1 + \ii \vp^2 &= 2 Z^0 \bar Z_1, \cr
  \vp^3             &= Z^1 \bar Z_1 -  Z^0 \bar Z_0; \cr
}$$
and the perturbation described above is  $\delta\vp^a(Z^B, \bar Z_B)$, where
$$\eqalign{
\delta\vp^1 + \ii\delta\vp^2 &= 2(\bar Z_1)^2 \Omega -2(Z^0)^2 \bar\Omega, \cr
\delta\vp^3 &= -2\bar Z_0 \bar Z_1 \Omega - 2Z^0Z^1 \bar\Omega, \cr
}$$
with $\Omega = Z^1\Theta^0 - Z^0\Theta^1$.
For the first few of these modes, the change $\delta E$ in energy is as
follows.
\item{(i)} If $\Theta^A = T^A$ constant, then
  $\delta E = 4\pi^2(T^A\bar T_A) (2/R-R)$; in other words, these modes are
  positive if $R<\sqrt{2}$ and negative if $R>\sqrt{2}$.  So the Hopf map
  is indeed unstable for $R>\sqrt{2}$.
\item{(ii)} If $\Theta^A = T^{AB} \bar Z_B$ with  $T^{AB}$ skew, then 
  $\delta E = 0$: this is a zero-mode.
\item{(iii)} If $\Theta^A = T^A_D Z^D$ with $T^A_A = 0$ (the trace part does
  not contribute to  $\delta\vp^a$) and $\bar T^A_B = T^A_B$, then
  $\delta E = 128\pi^2(T^B_A \bar T^A_B)/(3R)$: these modes are positive for
  all $R$.
\par\ni In conclusion, we can identify a perturbation of the Hopf map which
becomes a negative mode for  $R>\sqrt{2}$.  It seems likely that there are no
negative modes for  $R<\sqrt{2}$, \ie\ the Hopf map is then stable; but this
has not been proved.

\bn 4. {\scap Approximate Solutions on $S^3_R$ and $\rt$.}

\mn To begin with, let us investigate a one-parameter family of $Q=1$
configurations, which contains the Hopf map $\Phi: S^3_R \to S^2$, and also
contains one of the negative modes identified above.  This family of fields
$\Phi_\l$, with $\l$ a positive parameter, may be described geometrically
as follows.

Let $P$ denote a stereographic projection $P: S^3_R \to \rt$, and let $D_\l$
denote a dilation on $\rt$ with (constant) scale-factor $\l>0$.  Then
define $\Phi_\l$ by
$$
  \Phi_\l := \Phi \circ P^{-1} \circ D_\l \circ P\,.
$$
The energy of this field turns out to be
$$
  E_\l := E[\Phi_\l] = {64\pi^2\l R\over(\l +1)^2}
                        + {8\pi^2(\l^2+1)\over \l R}.
$$
Note that $E_1 = 16\pi^2(R+R^{-1})$, as must be the case
(since $\Phi_1 = \Phi$).

Now let us find the value of $\l$ for which $E_\l$ is a minimum.  A
straightforward calculation shows that if $R < \sqrt{2}$, then the minimum
occurs at $\l=1$ (\ie\ for the Hopf map);  whereas if $R > \sqrt{2}$, then the
minimum occurs when $\l$ is either of the roots of
$\l^2 + 2(1-\sqrt{2}R)\l + 1 = 0$.  In the latter case, the minimum value is
$E_\l({\rm min}) = 32\sqrt{2}\pi^2 - 16\pi^2/R$.  The soliton, instead of
being spread out over the whole of $S^3$, is then localized around a particular
point (the base-point of the  stereographic projection).

In the limit as $R\to\infty$, one gets a $Q=1$ configuration on $\rt$
which simply consists of an inverse stereographic projection $\rt\to S^3$
followed by the Hopf map (this kind of approximation has long been considered).
Its energy is $32\sqrt{2}\pi^2$, which is about 13\% higher than that of
the numerical solution.

As an example of higher-charge configurations, consider the field $\Psi_{m,n}$
which maps $(Z^0,Z^1)$ to the point with homogeneous coordinates
$\bigl[ (Z^0)^m/|Z^0|^{m-1}, (Z^1)^n/|Z^1|^{n-1} \bigr]$.  So $m=n=1$
gives the Hopf map.  For $m>1$ or $n>1$, the field $\Psi_{m,n}$ is
continuous but not smooth; however, its partial derivatives are continuous
and bounded on the complement of a set of measure zero in $S^3$, and so in
particular its energy is well-defined.  In fact, the charge and energy turn
out to be
$$\eqalignno{
  Q &= mn, \cr
  E[\Psi_{m,n}] &= 8\pi^2R + 4\pi^2(m^2+n^2)(R+2/R). &(5)\cr
}$$
If we minimize (5) with respect to the radius $R$, then we get
$$
  E_{{\rm min}} = 8\pi^2\sqrt{2(m^2+n^2)(2+m^2+n^2)}. \eqno(6)
$$
(The idea here is that higher-charge configurations prefer slightly more
living-space, and (6)
corresponds to the sphere-size in which they are most comfortable.)
The first few cases are as follows.
\item{(i)} If $m=2$ and $n=1$, then $Q=2$ and
  $E_{{\rm min}} = 32\pi^2 2^{3/4}\times1.24$,  \ie\ 24\% greater than
  our reference value.
\item{(ii)} If $m=3$ and $n=1$, then $Q=3$ and
  $E_{{\rm min}} = 32\pi^2 3^{3/4}\times1.70$, which is rather high.
\item{(iii)} If $m=n=2$, then $Q=4$ and
  $E_{{\rm min}} = 32\pi^2 4^{3/4}\times1.12$, quite close to the reference
  value.
\par\ni So the configuration $\Psi_{m,n}$ is a reasonable one.  But there
are several other simple expressions for higher-charge fields, for example
related to the ``rational-map'' ansatz for Skyrmions (cf.\ [3]), and these
deserve investigation as well.



\bn 5. {\scap Concluding Remarks.}

\mn Most recent work on Hopf solitons in the Skyrme-Faddeev system has
involved intensive numerical simulations.  The results reported in this
paper are aimed at complementing those studies.  One particular theme
is the way in which simple explicit solutions on $S^3_R$ become unstable
as $R$ increases, and collapse into localized structures.  Clearly there
is much scope for further investigation of such configurations.

Although the Hopf map is indeed a solution of the field equations, it
is not, strictly speaking, known to be stable for $R<\sqrt{2}$.  The
arguments for stability
of the identity map  in the Skyrme model [9, 10, 13] appear not to adapt
to this case, because of the nonlocality of the topological charge density
(and the absence of a Bogomolny-type lower bound on the energy).  It
would be useful if the inequality (1), (4) could be established;
in addition to providing a good lower bound on the energy, this would
prove that the Hopf map is stable for $R=1$.

\bn{\bf References.}
\item{[1]} Faddeev L and Niemi A J 1997 {\it Nature} {\bf387} 58--61
\item{[2]} Gladikowski J and Hellmund M 1997 {\it Phys.\ Rev.\ D} {\bf56}
           5194--9
\item{[3]} Battye R A and Sutcliffe P M 1997 {\it Solitonic Strings and Knots.}
           To appear in the CRM Series in Mathematical Physics
           (Springer-Verlag)
\item{} Battye R A and Sutcliffe P M 1998 {\it To be or knot to be?}
        hep-th/9808129.
\item{} Battye R A and Sutcliffe P M 1998 {\it Solitons, Links and Knots.}
        hep-th/9811077.
\item{[4]} Faddeev L and Niemi A J 1997 {\it Toroidal Configurations as Stable
           Solitons.} Preprint hep-th 9705176
\item{} Faddeev L and Niemi A J 1998 {\it Partially Dual Variables in SU(2)
           Yang-Mills Theory.} Preprint hep-th 9807069
\item{[5]} Hietarinta J and Salo P 1998 {\it Faddeev-Hopf knots: Dynamics of
           linked un-knots.} hep-th/9811053.
\item{[6]} Faddeev L 1975 {\it Quantisation of Solitons.} Preprint IAS
           Print-75-QS70, Princeton
\item{[7]} Kundu A 1986 {\it Phys.\ Lett.\ B} {\bf171} 67--70
\item{[8]} Kundu A and Rybakov Yu P 1982 {\it J.\ Phys. A} {\bf15} 269--75
\item{[9]} Manton N S and Ruback P J 1986 {\it Phys.\ Lett.\ B} {\bf181}
            137--40
\item{[10]} Manton N S 1987 {\it Comm.\ Math.\ Phys.} {\bf111} 469--78
\item{[11]} Jackson A D, Manton N S and Wirzba A 1989 {\it Nucl.\ Phys.\ A}
            {\bf495} 499--522
\item{[12]} Vakulenko A F and Kapitanskii L V 1979 {\it Sov.\ Phys.\ Dokl.}
          {\bf24} 433--4
\item{[13]} Loss M 1987 {\it Lett.\ Math.\ Phys.} {\bf14} 149--56


\bye